          [2001/04/25 1.1 (PWD)]
\documentclass[a4paper,twocolumn]{esapub2005} 
\usepackage{times}
\usepackage{natbib}
\usepackage{graphicx}

\title{The sky distribution of 511 keV positron annihilation line emission as
measured with INTEGRAL/SPI
}
\author{G. Weidenspointner}
\author{J. Kn\"odlseder}
\author{P. Jean}
\author{G.K. Skinner}
\author{P. von Ballmoos}
\author{J.-P. Roques}
\author{G. Vedrenne}
\affil{Centre d'Etude Spatiale des Rayonnements, BP 4346, 31028
Toulouse Cedex 4, France}
\author{\\P. Milne}
\affil{Steward Observatory, 933 N. Cherry Ave., Tucson, AZ, 85721, USA}
\author{B.J. Teegarden}
\affil{NASA/GSFC, LHEA, Greenbelt, MD 20771, USA}
\author{R. Diehl}
\author{A. Strong}
\affil{Max-Planck-Institut f\"ur extraterrestrische Physik, 85740
Garching, Germany}
\author{S. Schanne}
\author{B. Cordier}
\affil{DSM/DAPNIA/SAp, CEA Saclay, Gif-sur-Yvette, France}
\author{Ch. Winkler}
\affil{ESA/ESTEC, SCI-SD, 2201 AZ Noordwijk, The Netherlands}

\begin{document}


\maketitle

\begin{abstract}


The imaging spectrometer SPI on board ESA's INTEGRAL observatory
provides us with an unprecedented view of positron annihilation in our
Galaxy. The first sky maps in the 511~keV annihilation line and in the
positronium continuum from SPI showed a puzzling concentration of
annihilation radiation in the Galactic bulge region. By now, more than
twice as many INTEGRAL observations are available, offering new clues
to the origin of Galactic positrons. We present the current status of
our analyses of this augmented data set. We now detect significant
emission from outside the Galactic bulge region. The 511~keV line is
clearly detected from the Galactic disk; in addition, there is a
tantalizing hint at possible halo-like emission. The available data do
not yet permit to discern whether the emission around the bulge region
originates from a halo-like component or from a disk component that is
very extended in latitude.
%

\end{abstract}


\section{Introduction}
\label{intro}

The detection of 511~keV positron annihilation line emission from the
central region of our Galaxy was one of the early and important
successes of gamma-ray astronomy \citep{Johnson72,
Leventhal78}. Although positron annihilation gives rise to the
strongest gamma-ray line signal from our Galaxy, three decades of
dedicated observational and theoretical effort have fallen short of
unveiling the origin of Galactic positrons. A large variety of
potential astrophysical and exotic positron sources have been
proposed, including: interactions of cosmic rays in the interstellar
medium \citep{Ramaty70}; pulsars \citep{Sturrock71} and millisecond
pulsars \citep{Wang06}; the decay of radioactive nuclei expelled by
supernovae \citep{Clayton73}, novae \citep{Clayton_Hoyle74},
Wolf-Rayet stars
\citep{Prantzos_Casse86}, or hypernovae/gamma-ray bursts
\citep{Casse04}; compact objects harbouring black holes or neutron
stars \citep{Ramaty_Lingenfelter79}, low-mass X-ray binaries
\citep{Prantzos04}, and micro-quasars \citep{Guessoum06}; gamma-ray
bursts \citep{Lingenfelter_Hueter84};
pair production by gamma rays from so-called Small Mass Black Holes
with X-ray photons from the supermassive black hole at the center of
our Galaxy, Sgr~A$^\ast$ \citep{Titarchuk_Chardonnet06};
interactions of cosmic rays accelerated during episodes of mass
accretion onto the supermassive black hole Sgr~A$^\ast$
\citep{Cheng06, Totani06}; and decay or annihilation of (light) dark
matter \citep{Rudaz_Stecker88, Boehm04} or other exotic particles
and processes \citep[see e.g.\ references in ][]{Beacom_Yueksel06}.


Investigations of the
sky distribution of the annihilation radiation promise to provide
clues for the identification of the source(s) of positrons in our
Galaxy. Positrons may travel from their birth places before
annihilating, however, recent theoretical work suggests that positron
diffusion is limited \citep{Jean06a, Gillard06}. Thus the Galactic
distribution of annihilation radiation can be expected to resemble
that of the positron sources.

First maps of the annihilation radiation, limited to the inner regions
of our Galaxy, were obtained using the OSSE instrument on board the
Compton Gamma-Ray Observatory
in the 511~keV line and in positronium continuum emission
\citep[e.g.][]{Purcell97, Chen97, Milne00, Milne01a, Milne01b,
Milne02}. Some of these analyses also took advantage of data
obtained with the Gamma-Ray Spectrometer on board the Solar Maximum
Mission (SMM) and the TGRS spectrometer on board the Wind satellite.
Furthermore, the OSSE instrument allowed \citet{Kinzer99} and
\citet{Kinzer01} to study the one-dimensional distribution in
longitude and in latitude of diffuse emission, including
annihilation radiation, from the inner Galaxy.

With the commissioning of the imaging spectrometer SPI on board ESA's
INTEGRAL observatory, high spectral resolution mapping with improved
angular resolution has become feasible \citep{Jean03a,
Weidenspointner04, Knoedlseder05, Strong05, Weidenspointner06}.
Both the 511~keV line emission and the positronium continuum emission
were found to be brightest in a region of a few degrees around the
Galactic center; emission from the Galactic disk is much fainter,
implying that positron annihilation is concentrated in the central
regions of our Galaxy \citep{Knoedlseder05, Strong05,
Weidenspointner06}. In particular, the high bulge-to-disk flux ratio
of 1--3 (corresponding to a luminosity ratio of 3--9) measured in the
annihilation line imposes severe constraints on potential positron
sources. The most promising positron sources for the Galactic bulge
were found to be members of the old stellar population, specifically
Type~Ia supernovae (SN~Ia) and/or low-mass X-ray binaries
(LMXBs). Annihilation or decay of light dark matter was another
possiblilty. At least part of the faint emission from the Galactic
disk must be due to the $\beta^+$-decay of the radioisotopes $^{26}$Al
and $^{44}$Ti which are produced by massive members of the young
stellar population. However, none of these positron sources is without
caveats \citep[see e.g.][]{Knoedlseder05, Guessoum06, Schanne06}.

Additional insight into the origin, propagation, and annihilation of
positrons can be obtained from high-resolution spectroscopy. The
detailed shape of the annihilation line and the ratio of the fluxes
in the line and the positronium continuum depend on the physical
conditions of the interstellar medium in which positrons annihilate
\citep[see e.g.][]{Guessoum05}. Using observations of the Galactic
bulge, \citet{Jean06a} showed that positrons annihilate in the warm
phase of the interstellar medium. Recently, \citet{Beacom_Yueksel06}
pointed out that spectroscopy of Galactic emission above 511~keV can
be used to set an upper limit on the initial energy at which positrons
are injected into the interstellar medium \citep[see also][]{Sizun06}.


In this contribution, we present the current status of our studies
of the large-scale distribution of Galactic 511~keV annihilation
line radiation using more than 2 years of observations with the
spectrometer SPI. An update on SPI spectroscopy of positron
annihilation radiation is given in a companion paper by
\citet{Jean06b}. After describing our analysis methods, we first
present an updated sky map in the 511~keV line. We then summarize
our current results regarding the distribution of the bright 511~keV
line emission from the Galactic center region and regarding the
existence and distribution of more extended emission from the
Galactic disk and beyond. 



\section{Instrument and data analysis}
\label{data_analysis}

The SPI imaging spectrometer consists of an array of 19 actively
cooled high resolution Ge detectors,
which cover an energy range of 20--8000~keV with an energy resolution
of about 2.1~keV
FHWM at 511~keV. SPI
employs an active anti-coincidence shield made of bismuth germanate
(BGO), which also acts as a collimator. In addition to its
spectroscopic capabilities, SPI can image the sky with moderate spatial
resolution of about 3$^\circ$ FWHM using a tungsten coded aperture
mask. The fully coded field-of-view of the instrument is about
$16^\circ$.
A detailed description of the instrument was given by
\citet{Vedrenne03}.


\begin{figure}
\centering
\includegraphics[width=7.75cm,bbllx=24pt,bblly=285pt,bburx=471pt,bbury=508pt,clip=]{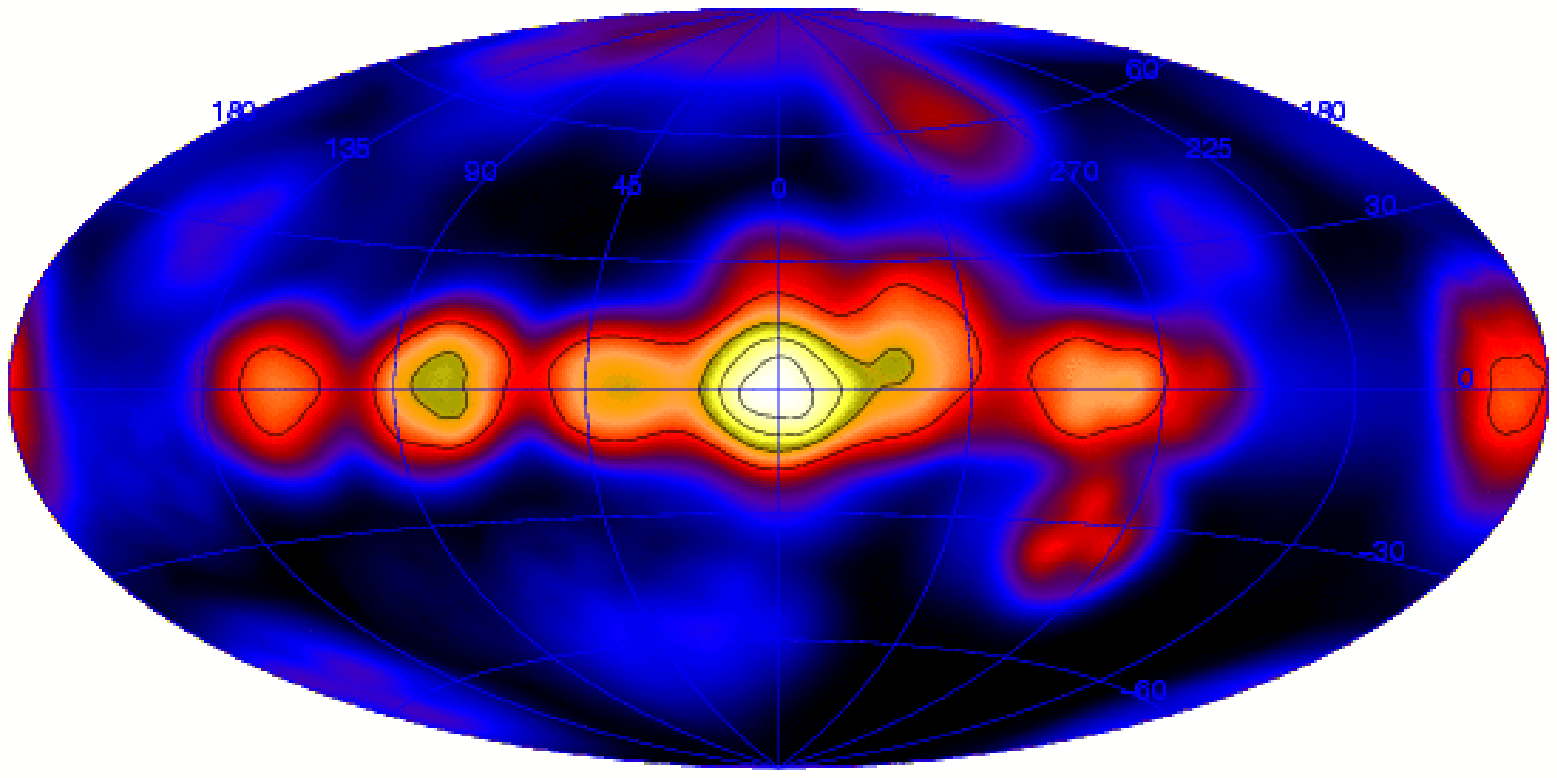}
\includegraphics[width=7.75cm,bbllx=24pt,bblly=285pt,bburx=471pt,bbury=508pt,clip=]{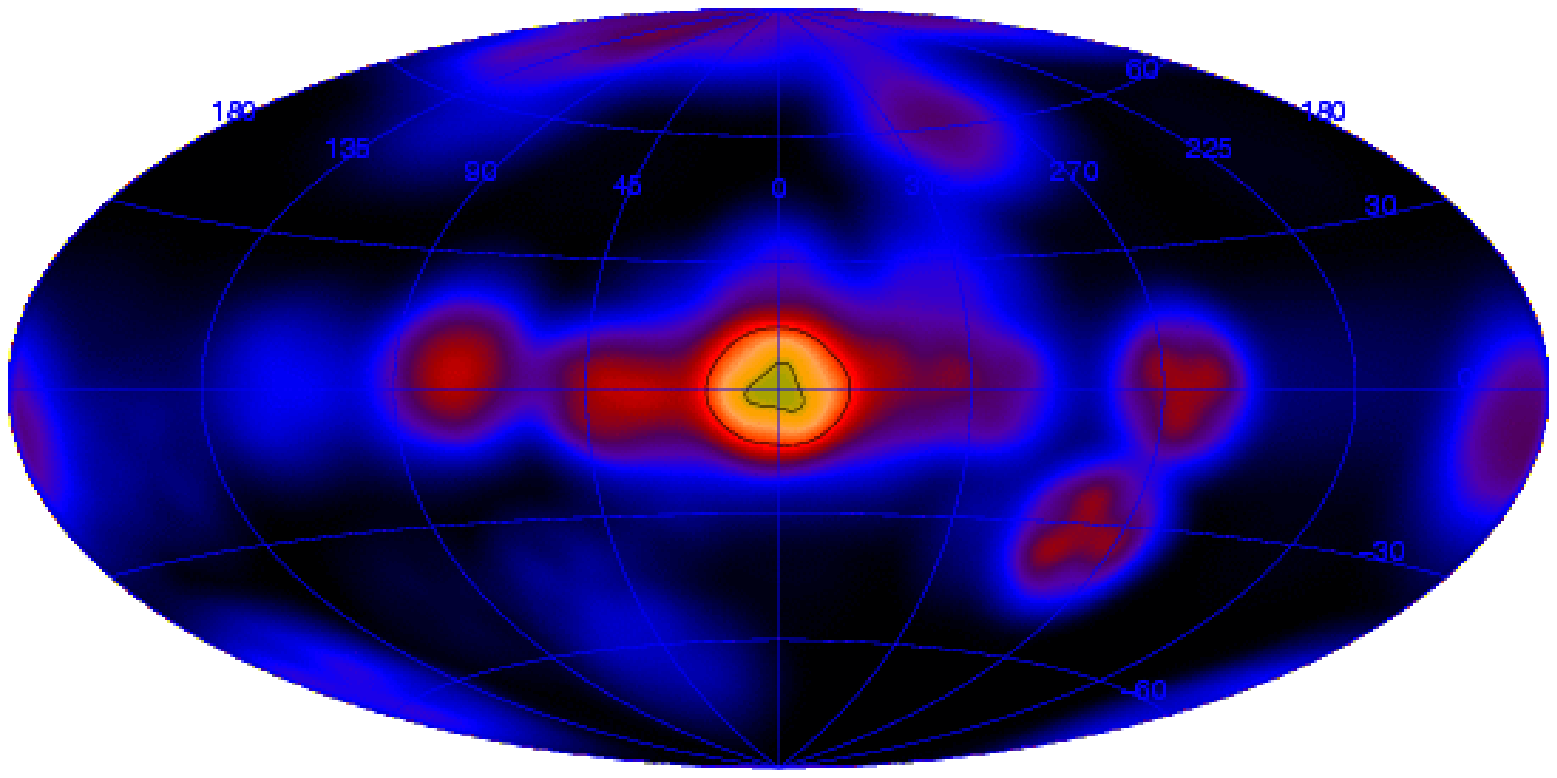}
\caption{Top panel: the exposure to the sky of the data set used in
this analysis. Contour levels are at $1$, $2$, $3$, and $4 \times
10^7$~cm$^2$~s. Bottom panel: the exposure to the sky after the first
year of the mission, depicted with the same color coding than
above. Contour levels are at $1$ and $2 \times
10^7$~cm$^2$~s.\label{exposure}}
\end{figure}

The results presented here are based on a data set that comprises
more than twice as many observations as that we used in our studies
after the first year of the mission \citep{Knoedlseder05, Jean06a,
Weidenspointner06}. Specifically, we have used the April 20, 2006
public INTEGRAL data release (i.e.\ three-day orbital revolutions
16-269, 277, 278, 283-285). These public data were supplemented by
instrument team observations of the Galactic center region and of
the Galactic plane and by recent private observations of SNR~1006,
Cen~X-4, and Sco~X-1 up to revolution 423. The observations were
taken during the epoch November 23, 2002 through April 1, 2006.
During this time a few very strong (notably during October 2003 and
September 2005), and several smaller, solar flares occurred. These
resulted in strong transient enhancements of the instrumental
background at 511~keV.  These exceptional periods were removed,
resulting in a cleaned data set that
consists of 18101 pointings
with a combined live time of $3.6 \times 10^7$~s.

The resulting exposure to the sky is depicted in the top panel of
Fig.~\ref{exposure}; for comparison, the exposure from the
observations of the first year of the mission is shown in the bottom
panel. To a first approximation, the exposure of the current data
set is about twice that achieved after the first year.
Of particular importance for the analyses presented here is the fact
that the exposure along the Galactic plane not only deepened, but
also became more homogeneous within about $|l| < 90^\circ$.
However, unfortunately the substantial inhomogeneity outside that
region
has not been alleviated. The exposure is still deepest around the
central region of our Galaxy, and most of the remaining, lower,
exposure away from that region is still concentrated in a rather
narrow band of $\pm 10^\circ$ along the inner part of the Galactic
disk. Investigations of the large-scale distribution of 511~keV line
annihilation therefore still face the dilemma that regions of
putative faint and extended emission remain poorly exposed. 
Specifically, the current inhomogeneity of exposure severely limits
our ability to distinguish between models for low surface brightness
emission outside the bulge region.  This is particularly pertinent to
investigating a potential halo distribution around the Galactic bulge,
and to determining the latitude distribution of the disk emission.

\begin{figure*}[t]
\centering
\includegraphics[width=13cm,bbllx=24pt,bblly=285pt,bburx=471pt,bbury=508pt,clip=]{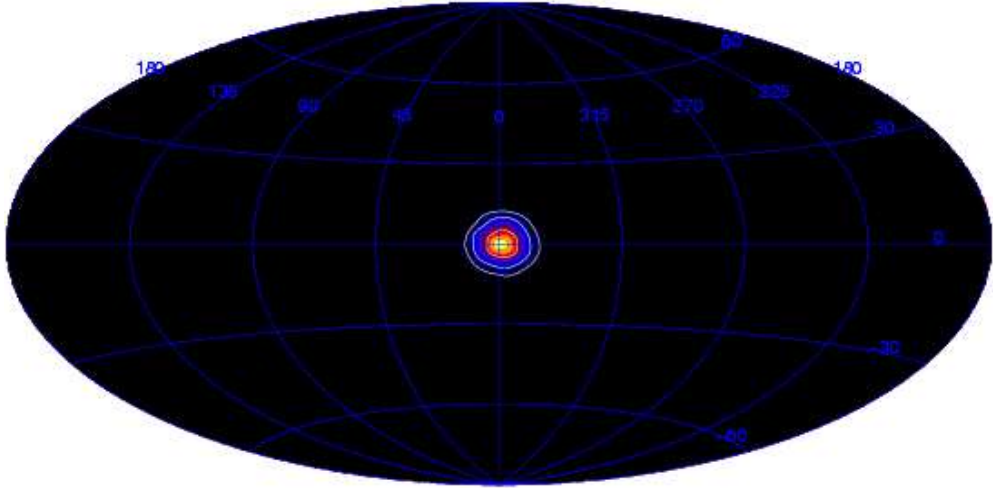}
\caption{An MREM sky map of the 511~keV positron annihilation line
emission. The contours indicate intensity levels of $10^{-2}$,
$10^{-3}$, and $10^{-4}$~ph~cm$^{-2}$~s$^{-1}$~sr$^{-1}$. Details are
given in the text.\label{511_map}}
\end{figure*}

In our analyses, we followed the procedures for modelling the
instrumental background that we established earlier \citep[for more
details, see e.g.][]{Knoedlseder05}.
%
%
The count rate in the detectors is the sum of time variable
instrumental background and signal
from celestial sources. The latter is variable
in time because even if the celestial source is intrinsically stable 
SPI's exposure to it varies in time as the instrument performs a
series of observations.
Source components have therefore to be distinguished from the dominant
background components by taking advantage of their differing time
variations.
Hence our approach for extracting the
source components consists of fitting time series of detector count
rates by a linear combination of so-called background templates
(time series of background components, explained below) and mask
modulated signals.

We performed our analyses in a 5~keV wide energy interval centered at
511~keV\footnote{When varying the width of the analysis energy
interval, the bulge flux varies as expected from the spectral results
by \citet{Jean06a}. Based on their two-Gaussian model for the 511~keV
line spectrum, the flux determined in our 5~keV analysis interval
corresponds to 82\% of the total line flux. The variation of the disk
flux is consistent with that of the bulge flux; detailed spectroscopy
results from the current data set are reported in \citet{Jean06b}.}.
In this energy range the instrumental background consists of two
components: the instrumental 511~keV background line, and an
underlying continuum background \citep{Jean03b,
Weidenspointner03,Teegarden04}. The variations in time of these two
background components differ, so they are modelled
independently \citep{Knoedlseder05}. The time variation of the 511~keV
background line is modelled using a linear combination of three
templates, the rate of saturating ($>8$~MeV) events in the Ge
detectors (GeDsat), a constant, and a linear increase in time. In this
context, a template represents a specific time series of background
count rates in a given detector.
The role of the first two templates is to account for
prompt and short-lived background components, the third template
accounts for long-lived background components.
The time variation of the continuum background underlying the
instrumental 511~keV line is modelled by assuming it is identical (per
unit energy) to that observed in an adjacent energy interval, for
which we chose 523--545~keV. To reduce statistical uncertainties, the
continuum time variation was smoothed \citep[see][]{Knoedlseder05}.

When applying this 4-component background model, the normalizations of
the three line background components are adjusted independently for
each detector to account for detector-to-detector variations of the
instrumental background. In addition, the normalizations of the GeDsat
component are adjusted for each orbital revolution, as in model {\tt
ORBIT-DETE} described in \citet{Knoedlseder05}. An additional
complication arises from the fact that during the epoch covered by our
data set, two SPI detectors failed (detector 2 on Dec.\ 7, 2003, or
IJD~1435, and detector 17 on Jul.\ 17, 2004, or IJD~1659). Each time a
detector fails, the detector-to-detector ratio of the instrumental
background changes significantly, especially in the 511~keV
background line and for detectors next to the failed detector. We
therefore split the data set into three periods, each corresponding to
a stable detector configuration, and simultaneously apply three
separate {\tt ORBIT-DETE} models.

\begin{table}[ht]
\begin{center}
\begin{tabular}[h]{ccc} \\ \hline \hline
Model Component & Flux [ph~cm$^{-2}$~s$^{-1}$] & $\lambda$ \\ \hline
\hline
\multicolumn{2}{c}{2 Nested Shells} & 1273.8 \\ \hline
0--0.5~kpc Shell & $(3.88 \pm 0.34) \times 10^{-4}$ & \\
0.5-1.5~kpc Shell & $(5.72 \pm 0.47) \times 10^{-4}$ & \\ \hline
\multicolumn{2}{c}{3 Nested Shells} & 1294.7 \\ \hline
0--0.5~kpc Shell & $(3.88 \pm 0.34) \times 10^{-4}$ & \\
0.5-1.5~kpc Shell & $(4.64 \pm 0.52) \times 10^{-4}$ & \\
1.5--5~kpc Shell & $(8.05 \pm 1.76) \times 10^{-4}$ & \\ \hline
\multicolumn{2}{c}{4 Nested Shells} & 1302.1 \\ \hline
0--0.5~kpc Shell & $(3.88 \pm 0.34) \times 10^{-4}$ & \\
0.5-1.5~kpc Shell & $(4.92 \pm 0.53) \times 10^{-4}$ & \\
1.5--5~kpc Shell & $(4.73 \pm 1.23) \times 10^{-4}$ & \\
5--8~kpc Shell & $(13.6 \pm 5.04) \times 10^{-4}$ & \\ \hline \hline
\multicolumn{2}{c}{2 Nested Shells and Young Disk} & 1310.7 \\ \hline
0--0.5~kpc Shell & $(3.62 \pm 0.34) \times 10^{-4}$ & \\
0.5--1.5~kpc Shell & $(4.74 \pm 0.49) \times 10^{-4}$ & \\
Young Disk & $(1.11 \pm 0.18) \times 10^{-3}$ & \\ \hline
\multicolumn{2}{c}{3 Nested Shells and Young Disk} & 1318.2 \\ \hline
0--0.5~kpc Shell & $(3.66 \pm 0.34) \times 10^{-4}$ & \\
0.5--1.5~kpc Shell & $(4.21 \pm 0.53) \times 10^{-4}$ & \\
1.5--5~kpc Shell & $(5.09 \pm 1.86) \times 10^{-4}$ & \\
Young Disk & $(0.94 \pm 0.19) \times 10^{-3}$ & \\ \hline
\multicolumn{2}{c}{4 Nested Shells and Young Disk} & 1319.3 \\ \hline
0--0.5~kpc Shell & $(3.68 \pm 0.34) \times 10^{-4}$ & \\
0.5--1.5~kpc Shell & $(4.36 \pm 0.55) \times 10^{-4}$ & \\
1.5--5~kpc Shell & $(4.86 \pm 1.87) \times 10^{-4}$ & \\
5--8~kpc Shell & $(5.82 \pm 5.34) \times 10^{-4}$ & \\
Young Disk & $(0.86 \pm 0.21) \times 10^{-3}$ & \\ \hline \hline
\multicolumn{2}{c}{2 Nested Shells and Old Disk} & 1312.4 \\ \hline
0--0.5~kpc Shell & $(3.72 \pm 0.34) \times 10^{-4}$ & \\
0.5--1.5~kpc Shell & $(4.33 \pm 0.52) \times 10^{-4}$ & \\
Old Disk & $(1.82 \pm 0.29) \times 10^{-4}$ & \\ \hline
\multicolumn{2}{c}{3 Nested Shells and Old Disk} & 1317.4 \\ \hline
0--0.5~kpc Shell & $(3.76 \pm 0.34) \times 10^{-4}$ & \\
0.5--1.5~kpc Shell & $(3.97 \pm 0.54) \times 10^{-4}$ & \\
1.5--5~kpc Shell & $(4.33 \pm 1.93) \times 10^{-4}$ & \\
Old Disk & $(1.52 \pm 0.32) \times 10^{-3}$ & \\ \hline
\multicolumn{2}{c}{4 Nested Shells and Old Disk} & 1318.0 \\ \hline
0--0.5~kpc Shell & $(3.75 \pm 0.34) \times 10^{-4}$ & \\
0.5--1.5~kpc Shell & $(4.11 \pm 0.57) \times 10^{-4}$ & \\
1.5--5~kpc Shell &  $(4.26 \pm 1.93) \times 10^{-4}$ & \\
5--8~kpc Shell & $(4.35 \pm 5.55) \times 10^{-4}$ & \\
Old Disk & $(1.41 \pm 0.35) \times 10^{-3}$ & \\ \hline
\end{tabular}
\caption{The results of fits in which bulge and/or halo emission was
modelled by nested shells of homogeneous emissivity, and disk emission
was described by either the young or the old stellar disk model of
\citet{Robin03}. $\lambda$ is the maximum likelihood ratio of the fits.
\label{explor_table}}
\end{center}
\end{table}


\section{Results}
\label{results}

\subsection{Imaging}
\label{results_imaging}

To obtain a model independent sky map of the 511~keV positron
annihilation line radiation, we employed the Multi-Resolution
Expectation Maximization (MREM) algorithm described in
\citet{Knoedlseder06}.  This algorithm is an extension of the
implementation of the Richardson-Lucy (RL) algorithm that we applied
in earlier analyses \citep[e.g.][]{Knoedlseder05}.

The resulting sky map is depicted in Fig.~\ref{511_map}. As in
earlier SPI analyses of the positron annihilation line
\citep{Knoedlseder05} and the positronium continuum
\citep{Weidenspointner06}, the only prominent 511~keV line signal is
that seen from the central region of our Galaxy. Any emission from
other sky regions is much fainter. We are investigating the effect of
the noise filtering threshold on MREM maps, as well as the effect of
smoothing scales on RL maps, in an effort to determine the best
approach for uncovering faint and extended annihilation emission that
we detect by model fitting (see Sec.~\ref{results_model_fitting}).

As with the first year observations, we find again that the centroid
of the emission is close to the Galactic center.  The dominant
emission from the central Galaxy appears to be more concentrated than
seen in the first year observations. This conclusion is confirmed by model
fitting (see Sec.~\ref{results_model_fitting}).
The total flux in the map is about $1.0 \times
10^{-3}$~ph~cm$^{-2}$~s$^{-1}$.

\subsection{Model Fitting}
\label{results_model_fitting}

A more quantitative approach for studying the Galactic distribution
of the observed extended line emission is model fitting, which we
performed using a maximum likelihood multi-component fitting
algorithm described in \citet{Knoedlseder05}.
In the present work, our focus is on investigating whether there
is 511~keV line emission outside the central region of the Galaxy. In
particular, we are trying to characterize the emission from the
Galactic disk,
and to assess whether there is evidence for emission from a halo
component. In the first year data, disk emission was already
marginally detected, while a stellar halo component (comprising emission
peaking at the Galactic center and fainter emission extending far
beyond the bulge region) could not yet be discerned from pure bulge
models.

In order to investigate the possible existence of faint and extended
emission from outside the central region of the Galaxy we fitted
simple and flexible models for bulge and halo emissions in the absence or
presence of simple disk models. The bulge and halos emissions were
represented by nested spherical shells of homogeneous emissivity,
centered at the Galactic center. The disk emission was described
either by the young (0--0.15~Gyr) or the old (7--10~Gyr) stellar disk
models as derived by \citet{Robin03}. The fit results are summarized
in Table~\ref{explor_table}.

It is evident from Table~\ref{explor_table} that there is significant
511~keV line emission from outside the bulge region of the Galaxy. The
maximum likelihood ratio $\lambda$ for the ``2 nested shells'' bulge
model
is only 1273.8, much lower than for any other model tested. Adding a
halo component (i.e.\ shells extending farther from the Galactic
center) and/or a disk model significantly improves the fits. The disk
models are favoured over the halo models since either disk model
improves the fit more than additional shell/halo components. In any
combination of halo and disk components, the disk component is always
significantly detected.  When combined with a disk, the 1.5--5~kpc
shell is still marginally detected, which provides a tantalizing hint
at possible halo-like emission; the 5--8~kpc shell is then not
required. The fluxes from the two innermost shells, describing the
bulge region, are remarkably robust and independent of the presence of
other model components, reflecting the brightness of the bulge region
of our Galaxy in annihilation radiation.  For the two-shell ``bulge
only'' model, we obtain a total flux of $(0.96 \pm 0.06) \times
10^{-3}$ph~cm$^{-2}$~s$^{-1}$, in excellent agreement with
\citet{Knoedlseder05}. 
For all other models in Table~\ref{explor_table}, we obtain total
fluxes in the range (1.7--3.1)~$\times
10^{-3}$ph~cm$^{-2}$~s$^{-1}$. Considering only models including disk
components, but excluding models with four shells (since the large
flux attributed to the 5--8~kpc shell is very uncertain), we obtain
bulge+halo fluxes in the range (0.8--1.3)~$\times
10^{-3}$ph~cm$^{-2}$~s$^{-1}$ and disk fluxes in the range
(0.9--1.8)~$\times 10^{-3}$ph~cm$^{-2}$~s$^{-1}$. The bulge-to-disk
(B/D) flux ratio is found to be between 0.4 and 1.4.  These values are
lower than the range 1--3 determined by
\citet{Knoedlseder05} using the first year of SPI
observations. Although mostly consistent within statistical errors, we
tend to find now lower bulge fluxes and higher disk fluxes than with
the first year data. Our values for the B/D flux ratio lie within the
wide range of 0.2--3.3 obtained by \citet{Milne00} from OSSE/SMM/TGRS
observations. 

\begin{table*}[ht]
\begin{center}
\begin{tabular}[h]{cccc} \\ \hline
Bulge/Halo Model & Bulge/Halo Flux & Old Stellar Disk Flux &
$\lambda$
\\
                 & [ph~cm$^{-2}$~s$^{-1}$] & [ph~cm$^{-2}$~s$^{-1}$] &
\\ \hline
Dwek et al.\ (1995) stellar bulge E3 & $(8.67 \pm 0.38) \times
10^{-4}$ &
$(1.62 \pm 0.29) \times 10^{-3}$ & 1281.1 \\
Dwek et al.\ (1995) stellar bulge G3 & $(8.48 \pm 0.36) \times
10^{-4}$ &
$(1.60 \pm 0.29) \times 10^{-3}$ & 1295.4 \\
Freudenreich (1998) stellar bulge S & $(8.87 \pm 0.39) \times
10^{-4}$ &
$(1.57 \pm 0.30) \times 10^{-3}$ & 1262.7 \\
Freudenreich (1998) stellar bulge E & $(8.87 \pm 0.39) \times
10^{-4}$ &
$(1.56 \pm 0.30) \times 10^{-3}$ & 1264.1 \\
Robin et al.\ (2003) stellar halo & $(2.35 \pm 0.10) \times 10^{-3}$
&
$(1.16 \pm 0.30) \times 10^{-3}$ & 1319.6 \\
\end{tabular}
\caption{The results of fits to the 511~keV line data of the
preferred triaxial or bar-shaped stellar bulge models E3 and G3 by
\citet{Dwek95} and S and E by \citet{Freudenreich98}, and of the
stellar halo model by \citet{Robin03}. $\lambda$ is the maximum
likelihood ratio of the fits. \label{bulge_table}}
\end{center}
\end{table*}

We then investigated other characterizations of the bulge and disk
emission. In a first step, and to compare with previous work, we
modelled the bulge emission using an ellipsoidal distribution with a
Gaussian radial profile in longitude and latitude, with full-widths at
half maximum in longitude and latitude $\Gamma_l$ and $\Gamma_b$.
This bulge model was combined with the young stellar disk description
by \citet{Robin03}; its three parameters are the disk scale length
$h_{R_+}$, the scale length of the central disk hole $h_{R_-}$, and
the axis ratio $\epsilon$ \citep[see definitions in Table~3 of
][]{Robin03}. We find values of $\Gamma_l =
{6.5^\circ}^{+1.1^\circ}_{-0.9^\circ}$ and $\Gamma_b =
{5.1^\circ}^{+0.8^\circ}_{-0.8^\circ}$ for the bulge component. These
values indicate a bulge extent that is slightly smaller than the FWHM
of about $8^\circ$ inferred from the first year data
\citep{Knoedlseder05}. 
We note that \citet{Kinzer01} found values of $\Gamma_l = 6.3^\circ
\pm 1.5^\circ$ and $\Gamma_b = {4.9^\circ} \pm 0.7^\circ$ in an
analysis of OSSE observations, in very good
agreement with our measurement.
The parameter values for the disk component are $h_{R_+}
\sim 4$~kpc, $h_{R_-} \sim 3$~kpc, and $\epsilon \sim 0.3$. As will be
discussed below, none of these disk parameters is well constrained by
the current data (for comparison, the young stellar disk is defined by
$h_{R_+} = 5$~kpc, $h_{R_-} = 3$~kpc, and $\epsilon = 0.014$). Our
disk parameters correspond to a FWHM in longitude of about $70^\circ$
for the 511~keV line emission, which is about twice as large as
found by \citet{Kinzer01} with OSSE.  The maximum likelihood
ratio for this model is $\lambda = 1315.4$, and the bulge and
disk fluxes are $(7.04 \pm 0.32)
\times 10^{-4}$~ph~cm$^{-2}$~s$^{-1}$ and $(1.41 \pm 0.17) \times
10^{-3}$~ph~cm$^{-2}$~s$^{-1}$, respectively. In this case, the B/D
flux ratio is about 0.5. Considering the number of model parameters
that have been optimized, this description of the data using an
ellipsoidal bulge and and a very extended (in latitude) disk is not
statistically superior to the two-shell bulge models combined with a
stellar disk (compare Table~\ref{explor_table}).

Rather than changing the aspect ratio of the Gaussian, more
improvement is possible by changing the profile.  The bulge model can
be improved by using e.g.\ two Gaussian distributions instead of one
ellipsoidal distribution. The values for the FWHM of the two Gaussians
are ${2.1^\circ}^{+1.4^\circ}_{-1.5^\circ}$ and
${8.0^\circ}^{+2.3^\circ}_{-1.0^\circ}$, respectively. The value of
$\lambda$ rises to about 1325, indicating that the 511~keV emission
may be peaked or cusped at the Galactic center.  The fluxes in the
narrow and wide Gaussian and in the disk model are about $1.5 \times
10^{-4}$~ph~cm$^{-2}$~s$^{-1}$, $7.4 \times
10^{-4}$~ph~cm$^{-2}$~s$^{-1}$, and $1.3 \times
10^{-3}$~ph~cm$^{-2}$~s$^{-1}$, respectively. Now the B/D flux ratio
is about 0.7.

To further investigate the distribution of the dominant annihilation
radiation from the central region of the Galaxy, we also fitted to
the 511~keV data some astrophysical models, namely the preferred
triaxial or bar-shaped stellar bulge models E3 and G3 by
\citet{Dwek95} and S and E by \citet{Freudenreich98}, and the
stellar halo model by \citet{Robin03}, each combined with the old
stellar disk. The results are summarized in
Table~\ref{bulge_table}. The stellar bar models E3, G3, S, and E provide
rather poor descriptions of the emission from the bulge region,
compared to the simple shell models summarized in
Table~\ref{explor_table}, and compared to ellipsoidal or Gaussian
descriptions.
It therefore appears that the distribution of the 511~keV line
emission does not follow the stellar bulge. The stellar halo model of
\citet{Robin03}, however, provides a fit that is of similar quality
than that of the best analytical models. As will be discussed below
(see Sec.~\ref{discussion}), it is too early to conclude that indeed
halo emission around the Galactic center region exists. The good
agreement between data and stellar halo model may mainly be due to the
fact that it models well the peaked emission at the Galactic
center. The existence of a sharp peak in the emission is already
hinted at by the good fit provided by the two-Gaussian bulge model
described above.  For the stellar halo model (which includes bulge
emission), the total 511~keV line flux is about $2.35 \times
10^{-4}$~ph~cm$^{-2}$~s$^{-1}$, and the (bulge+halo)-to-disk flux
ratio is about 2. This is consistent with the earlier analysis of
\citet{Knoedlseder05} and also with estimates of halo fluxes by
OSSE/SMM/TGRS \citep{Milne00,Milne06}.

\begin{figure*}[t]
\centering
\includegraphics[width=5.25cm,bbllx=38pt,bblly=376pt,bburx=292pt,bbury=666pt,clip=]{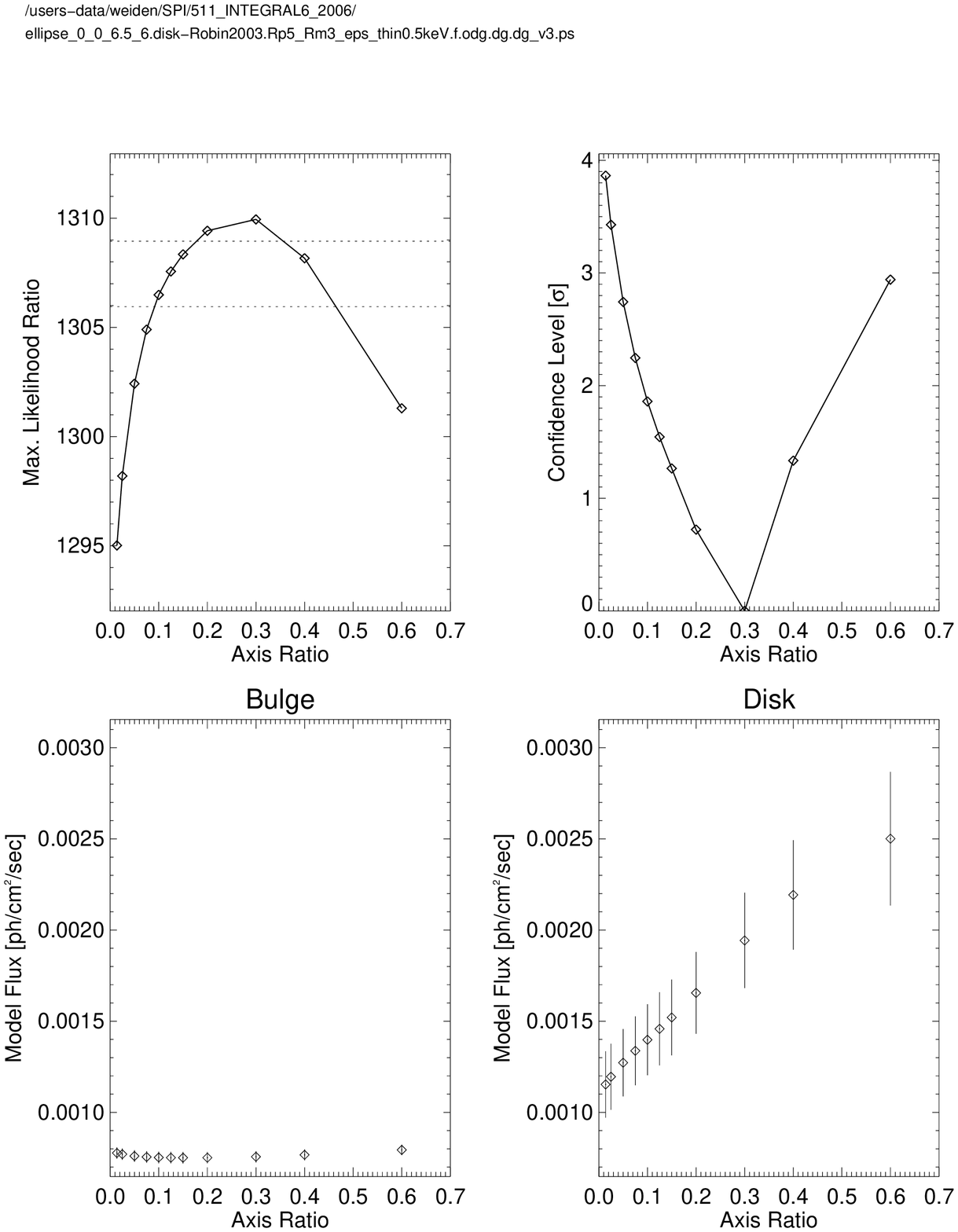}
\hspace*{0.25cm}
\includegraphics[width=5.25cm,bbllx=38pt,bblly=376pt,bburx=292pt,bbury=666pt,clip=]{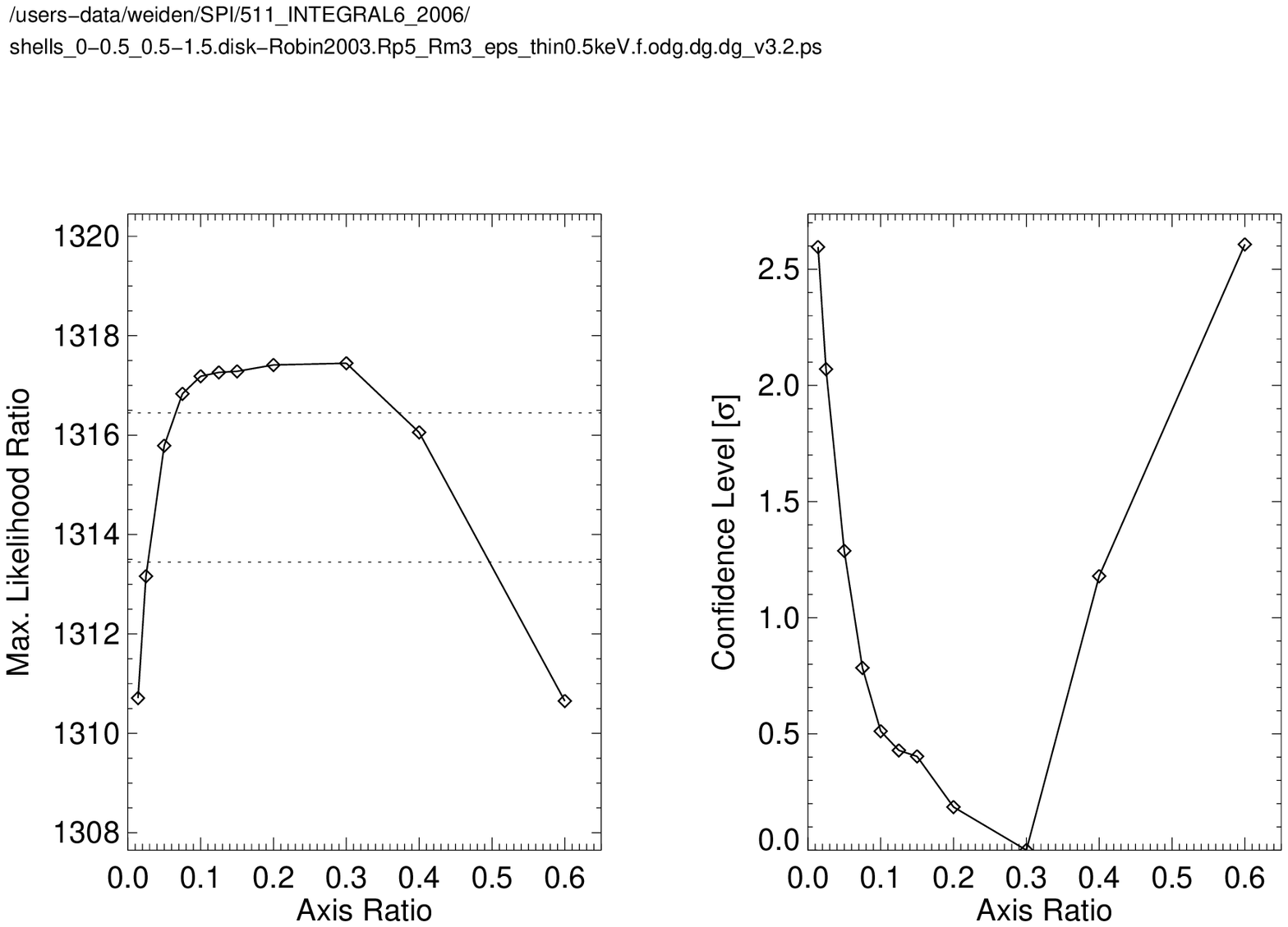}
\hspace*{0.25cm}
\includegraphics[width=5.25cm,bbllx=38pt,bblly=376pt,bburx=292pt,bbury=666pt,clip=]{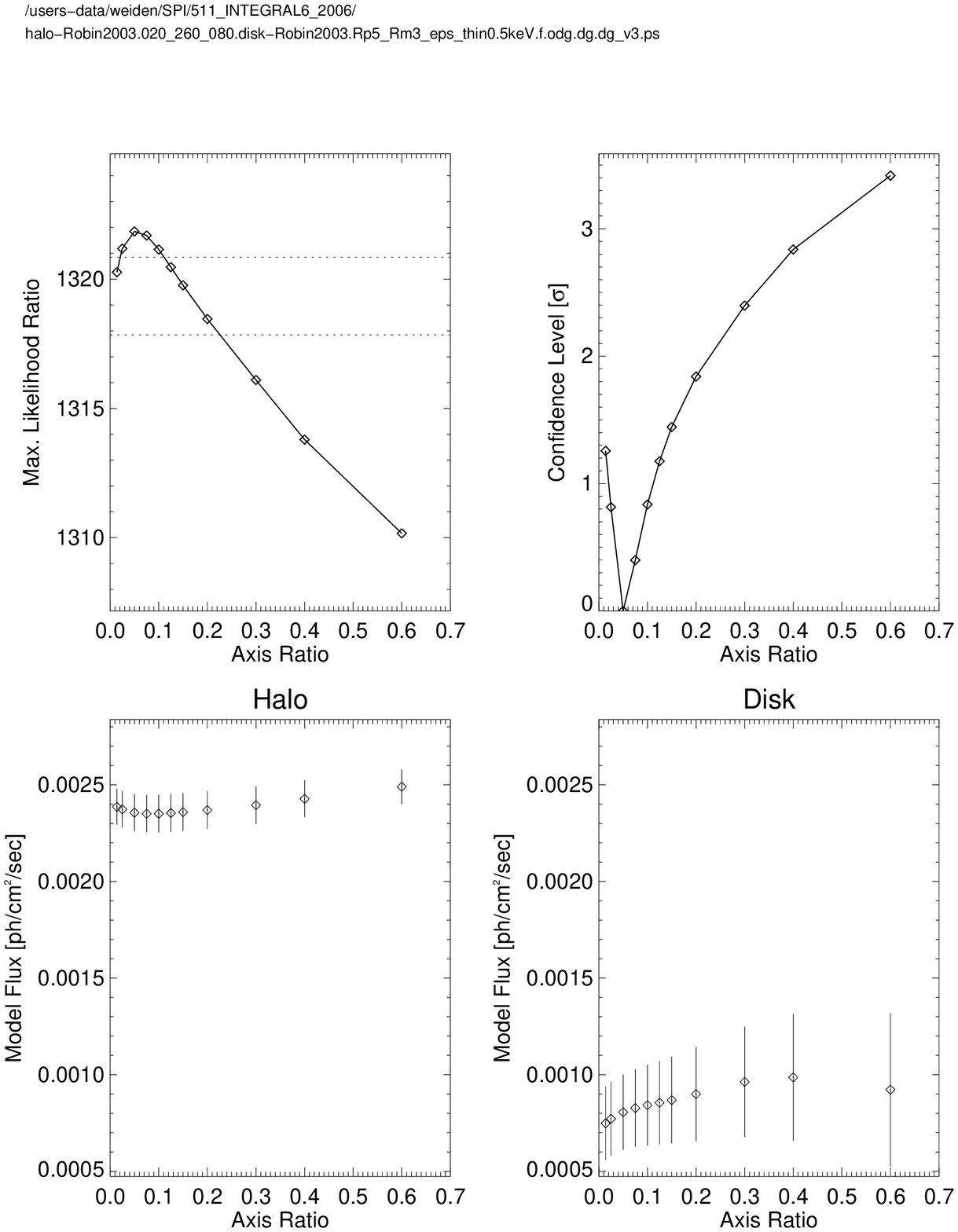}
\caption{Left panel: the variation of the maximum likelihood ratio $\lambda$
when fitting the data with the \citet{Robin03} young disk plus an
ellipsoidal model (FWHM $\Gamma_l = 6.5^\circ$ and $\Gamma_b =
6.0^\circ$) of the bulge emission. Center panel: as left panel, except
that the bulge emission is modelled by a combination of 0--0.5~kpc and
0.5--1.5~kpc homogeneous shells. Right panel: as left and center
panels, except that the emission from the central region of the Galaxy
is described by the stellar halo model of \citet{Robin03}. The dashed
lines indicate decreases of $\lambda$ by 1 and 4, corresponding to 1
and 2$\sigma$ confidence levels for 1 degree of freedom.
\label{axis_ratio}}
\end{figure*}

Our final step to date in investigating whether there is 511~keV line
emission outside the central region of the Galaxy, and whether it can
be uniquely described by rather simple models, was to vary the axis
ratio parameter in the \citet{Robin03} young and old disk
models. Combined with either the two-shell (see
Table~\ref{explor_table}), with an ellipsoidal ($\Gamma_l = 6.5^\circ$
and $\Gamma_b = 6.0^\circ$) description of the bulge emission, or with
the stellar halo model by \citet{Robin03}. As an example, for the
young disk the variation of the maximum likelihood ratio $\lambda$
with the axis ratio $\epsilon$ for the three descriptions of the
emission from the central region of the Galaxy are depicted in
Fig.~\ref{axis_ratio}. The results for the old stellar disk are
similar. We note that the axis ratios for the \citet{Robin03} young
(0--0.15~Gyr) and old (7--10~Gyr) stellar disks are 0.014 and 0.0791,
respectively.  For both bulge models, the preferred axis ratio is
about 0.3, much larger than the young disk value, in particular. The
uncertainty is, however, rather large.
For the stellar halo, the preferred value of $\epsilon$ is similar
to those for the \citet{Robin03} stellar disks, and the uncertainty is much
smaller.
These results clearly demonstrate that there is faint, extended
emission around the central region of the Galaxy, but with the
current data it is impossible to decide whether this emission is due
to a halo-like component, or a disk very extended in latitude.

As mentioned above, we have begun to study the longitude
distribution of the Galactic 511~keV line emission by determining
the best-fit values not only for the axis ratio $\epsilon$, but also
for the disk scale length $h_{R_+}$ and for the scale length of the
central hole $h_{R_-}$ for the \citet{Robin03} young stellar disk
parameterization. Their best fit parameters correspond to a
FWHM in longitude of about $70^\circ$ for the 511~keV line emission. 
We have also begun to study the longitude distribution by dividing the
disk emission into longitude bins. First results indicate that we do
not yet detect 511~keV emission from disk regions that are more than
$50^\circ$ in longtiude from the Galactic center.  Both results
indicate that 511~keV line emission is brightest from the inner
Galaxy.  More exposure, in particular in the outer Galaxy, will be
required to better determine the longitude profile.


\section{Summary and discussion}
\label{discussion}

The analyses we have performed to date, taking advantage of more
than 2 years of observations, clearly demonstrate that SPI detects
511~keV annihilation line emission from outside the central or bulge
region of the Galaxy. In similar, earlier analyses using only the
first year of observations, the disk was only marginally detected by
SPI, while the emission from the central region of the Galaxy could
statistically be equally well described by pure bulge or bulge and halo
distributions. Before the launch of INTEGRAL, positron annihilation
from the Galactic disk had already been detected using OSSE
observations, sometimes supplemented by SMM and TGRS data
\citep[e.g.][]{Milne00, Kinzer01}. However, constraining the spatial
characteristics of the disk emission, in particular its latitude
distribution, eluded OSSE/SMM/TGRS, as did a conclusive result
regarding the existence of halo emission.

While SPI has not yet solved these two major goals of positron
astronomy, the current status of our studies with SPI does represent
significant progress and promises much more. SPI now detects disk
emission at a significance level of about 6$\sigma$ (see e.g.\
Table~\ref{explor_table}), thus confirming the earlier detections with
the OSSE/SMM/TGRS instruments. In addition, with SPI we find
tantalizing hints at possible halo-like emission. When representing
the halo-like emission by extended shells as in
Table~\ref{explor_table}, we detect the 1.5--5~kpc shell at the
$\sim2.5\sigma$ level even when a disk model is included in the fit.
Another indication for potential halo emission is the good fit of the
emission from the central region of the Galaxy provided by the stellar
halo model of \citet{Robin03}, which comprises emission peaking at the
Galactic center and fainter emission extending far beyond the bulge
region.
Finally, the existence of faint, extended emission around the Galactic bulge
region is demonstrated by the fact that when a halo component is not
included in the model disks
with very wide latitude extents are favoured. When describing the
central emission with a stellar halo,  the
preferred latitude extent of the disk is much smaller (see
Fig.~\ref{axis_ratio}). 

Despite this tantalizing indications for halo-like emission, some
caution is in order 
because of the inhomogeneous exposure to the sky (see
Fig.~\ref{exposure}). With the current exposure to the sky, we cannot
exclude that most or all of the flux attributed to the 1.5--5~kpc
shell model is actually originating from the Galactic disk,
particularly in case the disk is extended in latitude. Similarly, the
stellar halo model by \citet{Robin03} may describe the emission from
the central region of the Galaxy well mainly because the model is
rather peaked or cusped at the Galactic center. The existence of such
a peak or cusp is indicated by the good fit provided by the
two-Gaussian (FWHM about $2.1^\circ$ and $8.0^\circ$) bulge
model. Since both the emission and the exposure are maximal around the
Galactic center, it is possible that the fits result is driven by a
rather small sky area in the bulge. Lower surface brightness emission
around the bulge would fall in sky regions that have
received relatively little exposure. 
They therefore carry comparatively little weight in the fitting. 

More observations around the Galactic bulge region 
will be needed to alleviate the existing systematic biases due to the
current inhomogeneity of the exposure. With the existing data, we
cannot yet discern whether the emission around the bulge region
originates from a halo-like component or from a disk component that is
very extended in latitude. In particular, the latitude distribution of
the disk cannot be determined independently of assumptions about the
distribution of emission in and around the bulge region. 
Thus the situation reached with the INTEGRAL data with more than 2
years of data already starts to improve on what was possible 
with the combination of data from OSSE/SMM/TGRS at the end of the 9 year CGRO
mission \citep{Milne00,Milne06}. It is hoped that INTEGRAL will continue to
operate for many years and make further advances possible.

The current data indicate that the disk is brightest in 511~keV line
emission in the inner Galaxy. Preliminary results indicate a FWHM in
longitude of about $70^\circ$ for the 511~keV line emission, which is
about twice as large as determined by \citet{Kinzer01} with OSSE. More
exposure, in particular in the outer Galaxy, will be required to
better determine the longitude profile.




Using the first year of SPI observations, the emission from the
central region of the Galaxy could be equally well described by
various astrophysical stellar bulge and halo distributions, or simple
analytical shapes \citep{Knoedlseder05}. With the current, increased
data set this is no longer true. The preferred stellar bulge models
derived by \citet{Dwek95} and \citet{Freudenreich98} from COBE
observations in the infra-red provide a poorer fit of the 511~keV line
distribution than simple analytical functions. It therefore appears
that the distribution of the 511~keV line emission does not follow the
stellar bulge. Of all astrophysical distributions that we tested, only
the stellar halo model by \citet{Robin03} could compete. However,
given the lack of an independent detection of genuine halo emission
from outside the bulge region and the inhomogeneity of the exposure,
it is possible that the halo model fares well mainly because it
provides a good description of the brightest emission close to the
Galactic center. The good quality of the halo model fit, and the
results from attempts at fitting the bulge emission with more than a
single ellipsoid, indicate that within a few degrees from the Galactic
center the annihilation emission exhibits some sort of ``peak'' or
``cusp''. More observations will be needed to further investigate
whether the emerging sub-structure of annihilation radiation from the
complex bulge region can be associated with gas and/or source
distributions.  The flux attributed to the stellar halo is consistent
with the earlier analysis of \citet{Knoedlseder05} and also with
estimates of halo fluxes by OSSE/SMM/TGRS \citep{Milne00,Milne06}.

The analyses reported here are ongoing; more refined results will be
presented elsewhere, including searches for point sources.  Looking at
the years ahead, reducing the inhomogeneity of the exposure of the
central region of the Galaxy is crucial for studying the large-scale
distribution of annihilation radiation. More observations around the
Galactic bulge region in particular, but also above and below the
disk, will be needed to alleviate the currently existing systematic
uncertainties.


\section*{acknowledgements}

Based on observations with INTEGRAL, an ESA project with instruments
and science data centre funded by ESA member states (especially the
PI countries: Denmark, France, Germany, Italy, Switzerland, Spain),
Czech Republic and Poland, and with the participation of Russia and
the USA.



\end{document}